\documentclass[11pt, a4paper]{article}
\usepackage[utf8]{inputenc}
\usepackage[T1]{fontenc}
\usepackage{newtxtext, newtxmath} 
\usepackage{microtype}            

\usepackage[margin=1.1in]{geometry} 
\usepackage{mathtools}
\usepackage{fancyhdr}
\usepackage{braket}
\usepackage{cite}
\usepackage{hyperref}
\usepackage{bbm}

\newcommand{\Tr}{\operatorname{Tr}}
\usepackage{titlesec}
\titleformat{\section}{\large\bfseries}{\thesection}{1em}{}
\titleformat{\subsection}{\normalsize\bfseries}{\thesubsection}{1em}{}
\usepackage[labelfont=bf, font=small]{caption} 
\usepackage{graphicx}
\usepackage{booktabs} 

\usepackage{authblk}

\usepackage{xcolor}
\hypersetup{
    colorlinks=true,
    linkcolor=black,
    citecolor=blue!70!black,
    urlcolor=blue!70!black
}

\begin{document}
\title{
Transport interpretation of entanglement Hamiltonian cumulants in integrable quantum quenches
}

\author[1]{Riccardo Travaglino}
\affil[1]{\textit{SISSA and INFN Sezione di Trieste, via Bonomea 265, I-34136 Trieste, Italy}}
\author[1]{Pasquale Calabrese}

\maketitle
\begin{abstract}
We study the dynamics of the cumulants of the entanglement Hamiltonian in interacting integrable models following global quantum quenches. Building on recent results based on space-time duality, we show that these cumulants are exactly given by the cumulants of currents of suitable conserved charges evaluated in the macrostate selected by the initial state. This establishes a direct connection between entanglement dynamics and transport, providing a transport counterpart to the quasiparticle picture that successfully describes the evolution of the von Neumann entropy. In the free-fermion and conformal limits, our results reduce to the difference between the current cumulants carried by right- and left-moving excitations, recovering previously known expressions. In interacting integrable models, where a decomposition into independent right and left movers is no longer meaningful at the operator level, a similar structure survives at the level of the entanglement spectrum, yielding a unified description of entanglement Hamiltonian fluctuations across free, conformal, and interacting integrable systems.

\end{abstract}

\section{Introduction}
The study of entanglement in quantum many-body systems has become one of the central themes of modern theoretical physics. Indeed, entanglement provides a unique point of contact between seemingly disparate fields, ranging from quantum information theory, quantum many-body physics, and statistical mechanics \cite{damico2008,Calabrese_2009_1,horodecki2009,Latorre_2009,arealaws,Laflorencie_2016} to high-energy physics and holography \cite{Ryu2006,Ryu_2006_aspects,CASINI2004142,Casini_2007,headrick2019lectures}.
In the context of the non-equilibrium dynamics of closed quantum many-body systems \cite{RevModPhys.83.863,DAlessio:2015qtq,Gogolin_2016,calabrese2016introduction}, the interplay between entanglement propagation and transport has attracted considerable attention. A major breakthrough came from the study of quantum quenches in conformal field theories (CFTs) \cite{quench1,quench2,Calabrese_2009,Cardy_2016}, where it was realised that, in many cases, the time evolution of the bipartite Rényi entropies of a subsystem $A$,
\begin{equation}
S_A^{(\alpha)} = \frac{1}{1-\alpha} \log \operatorname{Tr}_A \rho_A^\alpha,
\end{equation}
can be understood in terms of pairs of entangled quasiparticles emitted during the quench and subsequently propagating ballistically across the system, thereby spreading quantum correlations and entanglement.
This quasiparticle picture was subsequently extended to both free and interacting integrable systems \cite{fagotti2008evolution,Essler_2016,alba1,alba2,calabrese_ln}, where it provided a remarkably successful description of entanglement dynamics following quantum quenches. In free theories, its scope was further broadened to encompass a wide variety of entanglement-related quantities, often yielding predictions of striking accuracy. These include, among others, the entanglement negativity, full counting statistics of entanglement, symmetry-resolved entanglement measures, and operator entanglement \cite{coser2014entanglement,Dubail_2017,groha2018full,bertini2018entanglement,alba2019entanglement,alba2019quantum,Bastianello_2020,parez2021quasiparticle,parez2021exact,Murciano2022,carollo_alba2022,carollo2022,Turkeshi2022,turkeshi2023,ares2023entanglement,santalla2023,rath2023entanglement,travaglino2025,Travaglino2025measurements,fulgado2025,travaglino2026dissipative,dipasquale2026}.
From an intuitive standpoint, the quasiparticle picture can be understood as stating that the growth of entanglement is, to leading order, a purely transport-driven phenomenon: entangled quasiparticle pairs generated by the quench propagate ballistically through the system, spreading quantum correlations as they move.
Despite the remarkable success of the quasiparticle picture and the deep physical insight it provides, its scope has been called into question in recent years by the emerging framework of space–time duality (SD) \cite{klobas2021entanglement,bertini2022growth,bertini2022entanglement,bertini2023nonequilibrium,bertini2024dynamics,bertini2025exactly,travaglino2026spacetimeduality,travaglino2026dynamical}. Within this framework, it was shown that the quasiparticle picture does not, in general, correctly describe the dynamics of Rényi entropies in interacting integrable systems, exposing limitations that are absent in free theories. In particular, Ref.~\cite{bertini2022growth} demonstrated that the evolution of Rényi entropies cannot generally be interpreted in terms of pairs of entangled quasiparticles. The only exception is the replica limit $\alpha\to1$, corresponding to the von Neumann entropy, for which the quasiparticle picture is recovered. For $\alpha\neq1$, preserving such an interpretation requires introducing ad hoc assumptions with no clear physical justification, such as quasiparticle velocities that explicitly depend on the Rényi index. These results naturally raise the question of whether a general and physically transparent relation between entanglement growth and transport exists that remains valid in both free and interacting integrable systems. 

In this work, we address the problem from a different angle, shifting the emphasis from Rényi entropies to the cumulants of the entanglement Hamiltonian (EH) \cite{EH}, 
\begin{equation}
    \rho_A = \frac{e^{-K_A}}{\mathcal{Z}}, \hspace{0.2cm}\mathcal{Z}= \Tr_A e^{-K_A}.
\end{equation}
where $\rho_A$ is the reduced density matrix of a subsystem $A$. 
Although the entanglement Hamiltonian $K_A$ encodes the complete information contained in the reduced density matrix and is therefore, in general, a highly intricate object, its structure is well understood in a number of important cases. In particular, powerful insights have been obtained for area-law states from quantum field theory \cite{Bisognano:1975ih,Bisognano:1976za,Casini_2011,Cardy_2016}, while in free systems exact correlation-matrix methods allow its explicit determination even in out-of-equilibrium settings such as quantum quenches \cite{ehFF1,EHFF2,klich-2018,digiulio2019entanglement,Eisler_2019,Di_Giulio_2020,Mintchev_2021,rottoli2024entanglementHamiltoniansquasiparticlepicture,travaglino2024,Rottoli_2022,Bonsignori_2024,eisler2024,Rottoli_2024,Eisler_2025,BERNARD2025117185,Bonsignori2026}. 
In interacting systems, by contrast, progress has been largely confined to numerical studies in equilibrium \cite{Dalmonte_2018,toldin2018,Giudici_2018,zhu2020entanglement,Kokail_2021,Li_2024,yang_2026}, and a clear analytical understanding of the entanglement Hamiltonian, particularly in out-of-equilibrium settings, is still missing. In this work, we take a step toward filling this gap by investigating the quench dynamics of the cumulants of the entanglement Hamiltonian itself
\begin{equation}
    \braket{K_A^n}_c = \partial_\lambda^n \log \Tr_A e^{-\lambda K_A}\bigl|_{\lambda=0}.
\end{equation}
These quantities contain the same information as the Rényi entropies, since both depend exclusively on the entanglement spectrum. However, they provide a complementary perspective by allowing entanglement to be interpreted in purely thermodynamic terms: the cumulants of the entanglement Hamiltonian characterise fluctuations of the entanglement spectrum in precisely the same way that thermal cumulants characterise fluctuations of the energy spectrum in a conventional thermodynamic ensemble. The relation between these two families of entanglement measures is readily established as
\begin{equation}
     \braket{K_A}  =  S_{A}- \log \mathcal{Z}, \hspace{0.5cm }\braket{K_A^{n>1}}_c = (-1)^\alpha \partial_\alpha^{n}\left((1-\alpha) S_A^{(\alpha)}\right)_{\alpha=1}.
\label{eq:cumulantsdefinition}
\end{equation}
Contrarily to the Rényi entropies, which have been the subject of intensive investigation in chaotic and integrable systems, still little attention has been devoted to the cumulants of the EH, except for some studies regarding the second cumulant \cite{deboer2019,Arias2023,arias2023b}, known also as capacity of entanglement. 

The structure of the paper is the following: sections \ref{sec:setup} and \ref{sec:SD} introduce the main setup and provide brief reviews of integrability and of the space-time duality approach respectively. Sections \ref{sec:free} and \ref{sec:interacting} then proceed with the evaluation of the cumulants of the EH in free and interacting theories, highlighting the emergent common structure. Section \ref{sec:cft} is then devoted to the study of the conformal limit of the interacting solution, which can be used as a test of the validity of the previous results by comparing with the known CFT expressions. Finally, we drive our conclusions in section \ref{sec:concl}.

\section{Setup}
\label{sec:setup}
We consider an integrable lattice Hamiltonian $H$ in $(1+1)$ dimensions. Given the subtleties surrounding the very notion of quantum integrability \cite{Caux_2011}, we focus throughout on Bethe-ansatz-solvable models \cite{bethezur1931,takahashi1999book,Korepin_1993}, which are characterised by the presence of an extensive (in system size) number of local conserved operators $[H,Q_i]=0$. Typical examples include the XXZ chain, which is genuinely interacting,
\begin{equation}
    H_{\rm XXZ} = J \sum_i S^x_i S^x_{i+1} +  S^y_i S^y_{i+1} + \Delta  S^z_i S^z_{i+1}, 
\end{equation}
and the XX chain, which is obtained in the limit $\Delta\to 0$, and is solvable by free fermionic techniques by mapping it, through a Jordan-Wigner transformation, to the tight binding model 
\begin{equation}
\label{eq:hopping}
    H= J\sum_i c_i^\dag c_{i+1} + h.c.\,,
\end{equation}
where $\{c_i,c_i^\dag\}$ are spinless fermionic operators. Even for interacting theories, integrability gives the possibility to determine the spectrum exactly\cite{Yang,Korepin_1993} thanks to the presence of stable quasi-particle excitations which determine the eigenstates,  and thermodynamic properties of the system \cite{yangyang,Zamolodchikov1989,takahashi1999book,Mussardo:2010mgq} through the Thermodynamic Bethe Ansatz (TBA) framework. In this framework, one describes macrostates in the thermodynamic limit $\ket{\{\rho_{\rm t},\vartheta\}}$ through the total density of states $\rho_{\rm t}(\mu)$, and the occupation functions $\vartheta(\mu)$. Here and below, the rapidity variable $\mu$ represents a convenient parameterisation of the dispersion relation $(p(\mu),\varepsilon(\mu))$~\cite{orbach1958}. For example, in free models the rapidity is chosen as the momentum itself, while in relativistic systems the choice is typically the natural rapidity $\varepsilon(\mu) = m \cosh(\mu)$. In interacting lattice models such as the XXZ chain, the parametrization is typically more involved. 

While in free models densities and occupation functions are independent, in integrable interacting models the effect of the interactions can be encoded precisely in an intertwining of the two through a set of coupled integral equations known as Bethe-Takahashi equations \cite{takahashi1999book}.
In essence, the TBA allows to deal with partition functions of Generalised Gibbs Ensembles (GGE), $ \log \mathcal{Z} = \log \Tr[e^{-\sum_i\beta_i Q_i}] $, where $Q_i$ are the conserved charges of the model and the $\beta_i$ are generalised chemical potentials. Considering theories with a single quasiparticle species, this takes the form
\begin{equation}
\label{eq:TBAbasic}
   \begin{split}
       \log \mathcal{Z} &=  \ell_A\int \frac{d\mu}{2\pi} p'(\mu) \log(1+\eta^{-1}(\mu)) \\
        \log \eta(\mu)&= \nu(\mu)+\int \frac{d\mu}{2\pi} K(\mu',\mu) \log(1+\eta^{-1}(\mu))
   \end{split}
\end{equation}
where $ K(\mu',\mu) $ is the scattering kernel, obtained up to a constant as the logarithmic derivative of the S-matrix of the theory (which vanishes for free theories), and $\nu(\mu)$ is the driving term, which fully contains the state dependence, and in this case takes the form $\nu(\mu)=\sum_i\beta_iq_i(\mu)$, $q_i(\mu)$ being the one-particle eigenvalues of the conserved charges over the state. The $\eta$ function is related to the occupation functions as
\begin{equation}
    \eta(\mu) = \frac{1-\theta(\mu)}{\theta(\mu)}\iff \theta(\mu) = \frac{1}{1+\eta(\mu)}
    \label{eq:relationetatheta}
\end{equation}
and is introduced simply for notational convenience. 

The TBA solution essentially provides the \emph{exact} renormalisation (or dressing) of all physical quantities as a consequence of interactions. In fact, the dressing of the charge eigenvalues carried by each quasiparticle can be found as $ q_i^{dr}(\mu)= \partial_{\beta_i} \log \eta(\mu)$, which gives \cite{doyon2020lecture}
\begin{equation}
   q_i^{dr}(\mu)  = q_i(\mu)  - \int \frac{d\mu}{2\pi} K(\mu',\mu) \theta(\mu') q_i^{dr}(\mu').
\end{equation}
In particular, while in free theories the excitations propagate at a fixed velocity ${\rm v}(k)=\partial_k \varepsilon_k$, in interacting models the velocity itself is dressed by the interactions, through the relation
\begin{equation}
\label{eq:dressedvelocity}
    {\rm v}^{dr}(\mu)\rho_t(\mu) = \frac{\varepsilon'(\mu)}{2\pi} - \int d\mu' K(\mu',\mu) \theta(\mu){\rm v}^{dr} (\mu')\rho_t(\mu') .
\end{equation} 

In the quench scenario, we consider the dynamics which arises after initialising the system in some homogeneous short range entangled state $\ket{\psi_0}$ which is expressed as a superposition of an extensive number of eigenstates of the Hamiltonian to guarantee thermodynamic behaviour, and let evolve through $H$ as $\ket{\psi(t)} = e^{-iHt}\ket{\psi_0}$; the main objects of interest are then entanglement measures relative to a subsystem $A$ which is small compared to the total system. Even for integrable Hamiltonians, characterising such evolution is a formidable task, and typically one has to focus on the evaluation of simple quantities or simplifying regimes which capture the main features of the evolution whilst still being computable. In particular, one typically focuses on the ballistic regime, in which both $\ell_A$ and $t$ are much larger than the microscopic details of the system, but much smaller than the total size $L$ (to avoid revival effects); different physical regimes are then accessed by varying the parameter $\zeta=x/t$.
 Several methods have been introduced to deal with the dynamics of integrable systems at this scale, and these include 
 Generalised Hydrodynamics (GHD) \cite{castroalvaredo2016emergent,bertini2016transport,doyon2020lecture,ESSLER2023127572}, the Quench Action approach \cite{caux2013,Caux_2016} and finally Space-Time duality itself. In particular, the latter is the only one which allows to approach the study of entanglement measures beyond von Neumann entanglement entropy, and so our discussion will rely heavily on it.

\section{Space-time duality}
\label{sec:SD}
In this section we review the main ideas of the space-time duality approach to the evaluation of Rényi entropies in integrable quenches. For a more detailed general introduction, we refer to the review \cite{bertini2025exactly}. 
The starting point of the approach relies on the observation that the quench dynamics from a generic lowly entangled initial state exhibits two main regimes: for $t \ll \ell_A$, the Renyi entropies of a subsystem $A$ exhibit a linear growth in time, while for $t\gg \ell_A$ they saturate to a value extensive in $\ell_A$. The latter regime corresponds to a situation in which the state $\rho_A$ has relaxed to a GGE \cite{Essler_2016} of the form $\rho_A(\infty) \sim e^{-\sum_i \beta_i Q_{i}}$ expressed in terms of the conserved quantities of the model which are activated by the initial state.
In this regime, the Rényi entropies can be evaluated via TBA techniques, as they can be expressed as a difference of two partition functions \cite{alba_quench_action_2017,Alba_Renyi_2017,Mestyan_2018},
\begin{equation}
    \log\Tr[\rho_A^{\alpha}] = \log \Tr[e^{-\alpha \sum_i \beta_i Q_i}] - \alpha\log \Tr[e^{-\sum_i \beta_i Q_i}] =\log  \mathcal{Z}_\alpha - \alpha\log \mathcal{Z} 
\end{equation}
where $\mathcal{Z}_\alpha$ is the partition function of a state in which all chemical potentials $\beta_i$ are rescaled by a factor $\alpha$. Using the TBA solution \eqref{eq:TBAbasic}, the difference is evaluated as
 \begin{equation}
 \label{eq:entanglementdensity}
   \begin{split}
      \frac{S_A^{(\alpha)}(t\gg \ell_A) }{\ell_A} &\approx  \frac{1}{ (1-\alpha)}\int \frac{d\mu}{2\pi} p'(\mu) \, \log\left((1-\theta(\mu))^\alpha+ \frac{\theta(\mu)^\alpha}{\log x_\alpha(\mu)}\right), \\
        \log x_\alpha &= \int d\mu K(\mu',\mu)   \log\left((1-\theta(\mu))^\alpha+ \frac{\theta(\mu)^\alpha}{\log x_\alpha(\mu)}\right).
        \end{split}
   \end{equation}
   In order to approach the early time regime, on the other hand, the main insight of the SD is then that the rate of growth of entanglement $S^{(\alpha)}(t)/t$ in the regime $t\ll \ell_A$ can be mapped to the density of entanglement obtained through \eqref{eq:entanglementdensity} of a theory in which the roles of space and time are swapped. In practice, the swap is performed by the substitution $p(\mu) \leftrightarrow \varepsilon(\mu)$ (which is a real-time version of the mirror transformation performed, for example, in \cite{Zamolodchikov1989})  and by the requirement that the occupation functions in the two theories have the same functional forms. According to the discussion of \cite{travaglino2026spacetimeduality}, the theory in the swapped channel can then be solved by a careful application of TBA techniques, which has to take into account some subtleties arising from writing a meaningful definition for the density of states in the swapped theory. This gives
\begin{equation}
\label{eq:SDentropy}
\begin{split}
    \frac{ S_A^{(\alpha)}(t\ll \ell_A)}{2t} &\approx  \frac{1}{ (1-\alpha)}\int d\mu \frac{\varepsilon'(\mu)}{2\pi}s(\mu)\log\left((1-\theta(\mu))^\alpha + \frac{\theta(\mu)^\alpha}{y_\alpha^{s(\mu)}}\right), \\
     \log y_\alpha(\mu) &= \int d\mu' K(\mu',\mu) s(\mu)\log\left((1-\theta(\mu))^\alpha + \frac{\theta(\mu)^\alpha}{y_\alpha^{s(\mu)}}\right),
\end{split}
\end{equation}
where we have introduced $s(\mu) = \text{sign}(\varepsilon'(\mu))$, and the factor $2$ in the denominator accounts for the presence of two boundaries of the subsystem $A$. For notational convenience, we will write
\begin{equation}
\label{eq:Lfunction}
\mathcal{L}_\alpha(\mu) = s(\mu)\log\left((1-\theta(\mu))^\alpha + \frac{\theta(\mu)^\alpha}{y_\alpha^{s(\mu)}}\right).
\end{equation}
The main feature which makes the SD result \eqref{eq:SDentropy} so remarkable is that the solution in both regimes $t\gg \ell_A$ and $t\ll \ell_A$ is fully determined by TBA data, namely the scattering kernel, the bare energy and momentum and the occupation functions. The latter can be determined by the initial state through the Quench Action approach (and this is the typical approach in numerical evaluations), but are also formally determined by the GGE reached after relaxation \cite{Mossel_2012, Essler_2016}, through the general TBA expression \eqref{eq:TBAbasic}. Hence in this sense the SD reflects a feature which was valid also within the quasiparticle picture, namely that the asymptotic data determine the full evolution of the entropies. On the other hand, it is clear that the two predictions above violate the quasiparticle picture for $\alpha \neq 1$: this would require a solution of the form \cite{alba1}
\begin{equation}
    S^{(\alpha)}(t) \sim \int \frac{d\mu}{2\pi} \min(2|{\rm v}^{dr}(\mu)|t,\ell_A)\rho_t(\mu) s^{(\alpha)}(\mu)
\end{equation}
where $s^{(\alpha)}(\mu)$ is the contribution to the Rényi entropy of each pair; this form is however not present unless $\alpha=1$, since in that case both $\log y_1$ and $\log x_1$ vanish. 
An important exception to this breakdown is represented by free theories, in which $K(\mu',\mu)=0$ and therefore $\log y_\alpha$ and $\log x_\alpha$ vanish for all values of $\alpha$. In the next two sections, we will discuss how these features have clear counterparts in the features of the cumulants of the Entanglement Hamiltonian.

\section{The free case}\label{sec:free}
We begin by considering free theories described by the Hamiltonian \eqref{eq:hopping}, in which the full EH following a quantum quench was derived in \cite{rottoli2024entanglementHamiltoniansquasiparticlepicture} through an operator-based formulation of the quasi particle picture. This is constructed by considering coarse grained degrees of freedom over fluid cells of size $\Delta$, as \begin{equation}
    b_{x,k} = \frac{1}{\sqrt{\Delta}}\sum_{z=0}^{\Delta-1} e^{ikz} c_{x+z}^\dag,
\end{equation}
where $x=n\Delta$ labels coordinates of the fluid cells. For quenches from Gaussian initial states, the EH turns out to be dependent only on the velocity of excitations ${\rm v}(k)=\partial_k\varepsilon_k$ and the number operators $\hat{n}_{x,k} = b_{x,k}^\dag b_{x,k}$, which satisfy $\braket{\hat{n}_{x,k}}=\theta(k)$ where the average is taken on the time-evolving state and $\theta(k)$ is the occupation function fixed by the initial condition. In particular, given a subsystem $A=[0,\ell_A]$ we have in the two regimes (see \cite{travaglino2024} for details)
\begin{equation}
\label{eq:ehfree}
\begin{cases}
        \hat{K}_A(t\ll \ell_A) = \int_{0}^\infty \frac{dk}{2\pi}\int_0 ^{2{\rm v}(k)t} dx \,\log\eta(k) \,\hat{n}_{x,k} 
    + \int_{-\infty}^0 \frac{dk}{2\pi}\int_{\ell_A-2|{\rm v}(k)|t} ^{\ell_A} dx\,\log\eta(k) \,\hat{n}_{x,k} \vspace{0.2cm}
    \\ \hat{K}_A(t\gg \ell_A) =\int_0^{\ell_A}dx \int_{-\infty}^\infty \frac{dk}{2\pi} \,\log\eta(k) \,\hat{n}_{x,k} 
  
\end{cases}
\end{equation}
where we have introduced the quantity $\log \eta(k) = \log \frac{1-\theta(k)}{\theta(k)}$ in a TBA-inspired notation. In fact, as discussed above this corresponds to the pseudoenergy of the free TBA, which is equal to the driving term $\nu(k)$ which determines the GGE at long times; this always admits a decomposition in the space of (quasi)local conserved charges, as $\nu(k) = \sum_i \beta_i q_i(k)$ where $q_i(k)$ is the one-particle eigenvalue of the $i$-th conserved charge. As stressed in the previous section for the Rényi entropies, this same quantity appears in the entanglement Hamiltonian in both regimes, implying that the same data which can be fixed by the final state GGE also determines the short time dynamics. This does not imply that the dynamics is frozen; in fact, the time dependence in \eqref{eq:ehfree} enters in a natural hydrodynamic way, i.e. through a redistribution of such number operators.

Since in this case we have access to the full operator description of the EH, it is immediate to access the cumulants exploting the gaussianity of the state. At large times, the system is in a GGE (the second line of \eqref{eq:ehfree} is precisely a GGE expressed in terms of the coarse grained operators), hence immediately
\begin{equation}
    \braket{K_A^n}_c=   \braket{\left(\sum \beta_i Q_i\right)^n}^{\theta(\mu)}_c
\end{equation}
where the expectation value on the right can be taken through TBA expressions given the occupation functions $\theta(\mu)$. On the other hand, in the 
small time regime evaluation gives
\begin{equation}
    \braket{K_A} = 2t \int \frac{dk}{2\pi} \theta(k) \text{sign}(k){\rm v}(k) \log \eta(k)\, ;
\end{equation}
considering again that $\log \eta(k) = \nu(k) = \sum_i \beta_i q_i(k)$, we see that the integrals reproduce known expressions for the average currents of conserved quantities \cite{castroalvaredo2016emergent,bertini2016transport} in the macrostate identified by the filling functions $\theta(k)$, up to the factor $\text{sign}(k)$. This separates the current carried by right movers with positive momenta from left movers with negative momenta, thus dividing the currents in a right moving $J^i_R$ and a left moving $J^i_L$. Therefore we can express the result as
\begin{equation}
\label{eq:currentaverage}
     \braket{K_A} = 2t \sum_i \beta_i \braket{J_R^i - J_L^i}^{\theta(k)},
\end{equation}
where $\hat{J}^i$ is the current corresponding to the conserved charge $Q_i$, and we stress that the average appearing on the right is over the macrostate described by the filling functions $\theta(k)$, while the one on the left is taken over the state $\rho_A \sim e^{-K_A}$ itself.
Note that this expression is meaningful since in free theories the current operators admit an operatorial decomposition of the form 
\begin{equation}
\label{eq:decomposition_current}
    \hat{J}=\int d\mu\hat{j}(k) = \int dk  \, q(k) {\rm v}(k) \hat{n}(k) \Rightarrow \hat{J}_{R/L} = \int_{+/-} dk  \, q(k) {\rm v}(k) \hat{n}(k)  ,
\end{equation} 
and there is therefore a natural rapidity factorization: as we will discuss in the next section, this is not valid for interacting theories.  Note that, while the average gives no insight on \emph{where} the current operators are acting, this is evident from the operator expression \eqref{eq:ehfree}:  this contains a net flow of right movers at the left edge of the system, and a net flow of left movers at the right edge of the system. 
We can then move to the evaluation of higher cumulants; these exhibit a similar form as the average, as
\begin{equation}
    \braket{K_A^n}_c = 2t \int \frac{dk}{2\pi}  \mathcal{P}_n(\theta(k)) |{\rm v}(k)| (\nu(k))^n, \label{eq:highercumulantsfree}
\end{equation}
where the polynomials $\mathcal{P}_n(\theta(k))$ appear in the expansion in $\alpha$ of $\log[(1-\theta(k))^\alpha + \theta(k)^\alpha]$ together with the factor $(\nu(k))^n$, and are ubiquitous in the evaluation of charge cumulants in integrable models \cite{myers2020,bertini2023nonequilibrium},
\begin{equation}
    \mathcal{P}_n(\theta(k))(\nu(k))^n = \partial_\alpha^n \log[(1-\theta(k))^\alpha + \theta(k)^\alpha].
\end{equation}
In particular, \eqref{eq:highercumulantsfree} can be interpreted as cumulants of current operators in the macrostate identified by the filling functions $\theta(k)$. It is convenient to consider the global current operators $J_R = \sum_i \beta_i J_R^i$ and   $J_L = \sum_i \beta_i J_L^i$; we therefore have the identification 
\begin{equation}
     \braket{K_A^n}^{\rho_A}_c = 2t \braket{(J_R - J_L)^n}_c^{\theta(k)}\label{eq:generalcumulantsfree}.
\end{equation}
This relation provides a remarkable connection between the cumulants of the EH, from which one potentially is able to reconstruct the full entanglement spectrum, and the transport of conserved quantities within the system. Considering that in the case $n=2$ the polynomial reduces to $\mathcal{P}_2(\theta(k)) = \theta(k)(1-\theta(k))$, the result matches the result of \cite{Arias2023} for the capacity of entanglement in a free quench.

\section{Interacting integrable models}\label{sec:interacting}
In this section we discuss the extension of the result to interacting integrable theories which represents the main result of this paper. 
Unlike the free-fermion case, interacting models do not admit a simple operator-level description for the entanglement Hamiltonian analogous to that of Ref. \cite{rottoli2024entanglementHamiltoniansquasiparticlepicture}.
As a consequence, one has direct access only to the cumulants, through the known expressions for the Rényi entropies reviewed in Section \ref{sec:SD}. 
In order to lighten the notation, we focus on theories with a single particle species, although extending the analysis to multi-species models is entirely straightforward.
The result \eqref{eq:SDentropy} for the Rényi entropies reads
\begin{equation}
    (1-\alpha)S_A^{(\alpha)} = 2t \int d\mu \frac{\varepsilon'(\mu)}{2\pi}\mathcal{L}_\alpha(\mu).
\end{equation}
where the function $\mathcal{L}_\alpha(\mu)$ is defined in equation \eqref{eq:Lfunction}.
As discussed in section \ref{sec:SD}, the Rényi entropies in the stationary state can be expressed in thermodynamic terms through TBA partition functions as $\log \mathcal{Z}_\alpha - \alpha \log \mathcal{Z}$.
In the dynamical regime, a similar discussion can be performed by considering the entanglement Hamiltonian: since
\begin{equation}
    \log\Tr[\rho_A^{\alpha}(t)] = \log \Tr[e^{-\alpha K_A(t)}] - \alpha\log \Tr[e^{-K_A(t)}] =\log  \mathcal{Z}^K_\alpha - \alpha\log \mathcal{Z}^K
    \label{eq:divisiontimedependent}
\end{equation}
the entropy result \eqref{eq:SDentropy} can be interpreted as arising from the difference of two partition functions relative to the time-dependent EH itself. These partiction functions can be found by separating \eqref{eq:SDentropy} as the difference in \eqref{eq:divisiontimedependent}, which gives a result which is analogous to the stationary case,
\begin{equation}
    \log \mathcal{Z}^K_\alpha =2t \int d\mu \frac{|\varepsilon'(\mu)|}{2\pi}  \log (1+\eta_\alpha^{-1})
\end{equation}
where the occupation functions are determined by the TBA equations
\begin{equation}
\label{eq:dressingSD}
    \log\eta_\alpha =  \alpha\tilde \nu(\mu) + s(\mu)\int \frac{d\mu'}{2\pi} s(\mu')K(\mu',\mu) \log (1+\eta_\alpha^{-1})  
\end{equation}
where $\tilde{\nu}(\mu)$ is some driving term that cancels out in the evaluation of the Rényi entropies. This driving term can be related to the one characterising the GGE by the requirement that the occupation functions obtained by \eqref{eq:dressingSD} are the same to those at final time, see for instance \cite{bertini2024dynamics,travaglino2026spacetimeduality}. Since these occupation functions are fixed by the TBA equations \eqref{eq:TBAbasic},
their equality (i.e. the equality of $\eta_1$, which is the one related to the occupation functions in \eqref{eq:relationetatheta}) implies the relation, obtained by differentiating in the chemical potentials $\beta_i$,
\begin{equation}
   \left[s(\mu) \partial_{\beta_i} \tilde \nu(\mu)\right]^{dr} = s(\mu) q_i^{dr}(\mu).
\end{equation}
In the free case, this expression reduces to $\partial_{\beta_i} \tilde{\nu} = q_i$, which yields $\tilde{\nu} = \nu = \sum_i \beta_i q_i$. Consequently, the early- and late-time dynamics are governed by the exact same driving term, as observed in the previous section. In the interacting case, however, the non-trivial action of the dressing inevitably makes these two driving terms distinct. This offers a new physical perspective on the observation in Section \ref{sec:SD} that Rényi entropy predictions cannot be naively interpolated between the early- and late-time regimes: we now see this disconnect arises because the respective partition functions are determined by entirely different driving terms.

It is now straightforward to evaluate the average of the EH as $S(t) - \log \mathcal{Z}^K$, by making use of the dressed velocity of excitations in equation \eqref{eq:dressedvelocity}. Direct evaluation gives
\begin{equation}
\label{eq:averageinteracting}
    \braket{K_A} = 2t\int d\mu \rho_t(\mu) {\rm v}^{dr} (\mu)\theta(\mu) s(\mu) \tilde \nu(\mu) .
\end{equation}
Just as above for free theories, this takes the form of an average current in the GHD language \cite{castroalvaredo2016emergent}. Contrarily to the free case, however, the interpretation in terms of right and left moving currents is not possible for a general interacting theory: in fact, the possibility to write $\braket{J_R - J_L}$ is specific to free fermions, since the currents admit an operator decomposition of the form \eqref{eq:decomposition_current}.
In the interacting case, the presence of the dressing implies that all different rapidities are coupled in a highly non-linear fashion. Even at the level of expectation values \cite{borsi2020currents,Borsi_2021}, although the expression for the average currents exhibits a seemingly decoupled form $\braket{J}= \int d\mu \rho_t(\mu) {\rm v}^{dr}(\mu) \theta(\mu)q(\mu)$, the coupling between the rapidities is actually hidden in the definition of $\rho_t {\rm v}^{dr} $, which contains information about all excitations of the model. The impossibility to write an equation \eqref{eq:decomposition_current} is also obvious when considering that the dressed velocity of excitations is \emph{state dependent}, implying that even the direction is not dependent on rapidity alone, giving rise to quasiparticles having curved trajectories in space-time (see for example \cite{alba2019entanglement,travaglino2026spacetimeduality}).
It is therefore clear that \eqref{eq:averageinteracting} has to be interpreted in a different way: in particular, we consider the operator $Q_C$ whose one-particle eigenvalues over Bethe states are given by $s(\mu) \tilde \nu(\mu)$ .  In general, this will not be strictly local because of the sign factor, but is expected to have a power-law decay; moreover, this is by construction a conserved quantity, as it is diagonal in the Bethe basis. To this charge we can associate a current $J_C$; the result \eqref{eq:averageinteracting} is equivalent to the relation 
\begin{equation}
    \braket{K_A(t)} = 2t \braket{J_C}^{\theta(\mu)}
\end{equation}
where as above the average on the right is taken over the macrostate identified by the filling functions $\theta(\mu)$. It is interesting to note that signatures associated with current-like terms in the entanglement Hamiltonian were also observed numerically in the non-integrable system investigated in Ref.~\cite{zhu2020entanglement}. At first sight, this may appear surprising, since a homogeneous system cannot support a net current of conserved quantities such as energy or particle number. The key observation is that $Q_C$ is odd under spatial inversion, implying that the corresponding current operator is even. Its expectation value is therefore not constrained to vanish by translational or inversion symmetry, explaining the emergence of such terms even in homogeneous states. 

The discussion can be extended to higher-order cumulants, although the resulting expressions rapidly become too cumbersome to be of practical use. For example, the second cumulant can be obtained by differentiation and it reads
\begin{equation}
    \braket{K_A^2}_c = 2t \int d\mu |{\rm v}^{dr} (\mu)|\rho_t(\mu) \mathcal{P}_2(\theta(\mu))\left([s(\mu)\tilde \nu(\mu)]^{dr}\right)^2 =2t \braket{J_C ^2}_c^{\theta(\mu)}
\end{equation}
which again confirms the relation with the current operator \cite{myers2020}. Note again that the dressing acts on $s(\mu) \tilde \nu(\mu)$. As a consequence, any decomposition into independent right- and left-moving contributions is necessarily impossible.

To avoid evaluating all cumulants, as the dressing becomes soon too involved \cite{myers2020}, we can directly focus on the cumulant generating function. This is given by $(1-\alpha) S^{(\alpha)} = \log \mathcal{Z}_\alpha - \alpha \log \mathcal{Z}$, and therefore 
\begin{equation}
\begin{split}
    \log \mathcal{Z}_\alpha - \alpha \log \mathcal{Z} = 2t \int d \mu  \frac{|\varepsilon'(\mu)|}{2\pi} \left(  \log (1+\eta_\alpha^{-1}) - \alpha \log (1+\eta^{-1}) \right)   
\end{split}
\label{eq:generatingfunction}
\end{equation}
where by \eqref{eq:dressingSD} the eta functions of interest satisfy 
\begin{equation}
    \partial_\alpha \log\eta_\alpha = s(\mu)[s(\mu) \tilde \nu(\mu)]^{dr}. \label{eq:derivativesign} 
\end{equation}
The last two equations together imply that the generating function is precisely identical to the one of \cite{myers2020} for the cumulant generating function of the current corresponding to the charge with eigenvalues $s(\mu) \tilde \nu(\mu)$, in the regime in which the quench is homogeneous and therefore the sign of the dressed velocity coincides with $s(\mu)$ \cite{travaglino2026spacetimeduality}. This result holds for all cumulants beyond the first. The first cumulant requires special treatment, as the second term in Eq.~\eqref{eq:generatingfunction} also contributes. This causes no inconsistency, however, since Eq.~\eqref{eq:cumulantsdefinition} contains the additional term $-\log \mathcal{Z}$, whose contribution exactly cancels the latter.
Therefore we conclude that all cumulants of the entanglement Hamiltonian are given by the cumulants of the current $J_C$ through 
\begin{equation}
    \braket{K_A^n}_c^{\rho_A(t)} = 2t \braket{J^n_C}^{\theta(\mu)}_c.
    \label{eq:finalformula}
\end{equation}
This is the main result of this paper, expressed in its most general form. Conceptually, this is providing a novel relation between the spread of entanglement  and transport after a quantum quench, which relates entanglement directly to currents without passing from the notion of entangling quasiparticles.
As discussed in Sec.~\ref{sec:free}, the two notions are essentially equivalent in free theories. In the following section, we demonstrate that this equivalence extends to conformal field theories as well.

\section{Conformal limit } \label{sec:cft}
In this section, we show that our results correctly reproduce the known structure of the entanglement Hamiltonian cumulants in conformal field theory (CFT) upon taking the conformal limit. We assume that Eq.~\eqref{eq:finalformula} remains valid in this regime; this property was previously established at the operator level for free fermions~\cite{rottoli2024entanglementHamiltoniansquasiparticlepicture}, making it a natural expectation here. Under this assumption, we recover the interpretation of $J_C$ as the difference between the currents carried by right- and left-moving excitations, which become well-defined and completely decouple in the conformal limit. For a lattice quench protocol, conformal behaviour emerges in the limit of small parameter variation in a critical Hamiltonian~\cite{Pollmann2013}. In the XXZ chain, for instance, this corresponds to an anisotropy quench $\Delta \to \Delta + \delta$ with $\delta \to 0$ and $\Delta < 1$. Because the pre-quench ground state is a filled Fermi sea and the quench injects minimal energy, the only resulting excitations are particle-hole pairs propagating ballistically at the Fermi velocity, naturally yielding a conformal description. Specifically, the dynamics decouples into right- and left-movers associated with excitations around the two Fermi points $\pm \Lambda_F$, allowing to connect directly to the conformal limit of integrable quantum field theories discussed in Refs.~\cite{Zamolodchikov:1992zr,Bazhanov1996}. 

Taking into account a standard quantum quench from the smoothed boundary state
$|\psi_0\rangle=e^{-\tau_0 H}|B\rangle$,
the driving term assumes the thermal form~\cite{quench1,Calabrese_2009,Cardy_2016},
$\beta\,\varepsilon_k$. Consequently, the current $J_C$, which is associated with the full driving term (including the inverse temperature $\beta$), can be written as 
\begin{equation}
    J_C=\beta\left(J_{E,R}-J_{E,L}\right),
\end{equation}
where \(J_{E,R}\) and \(J_{E,L}\) denote the energy currents carried by the right- and left-moving excitations, respectively.
Given the independence between right and left movers which implies $\braket{J_R^p J_L^q}_c=0$, we have
\begin{equation}
     \braket{K_A^n}^{\rho_A}_c = 2t \beta^n \left(\braket{J^n_{E,R}}_c ^{\beta}+(-1)^n\braket{J_{E,L}^n}_c^{\beta}\right)
     \label{eq:ehcft}
\end{equation}
where the averages of the currents are over thermal CFT fixed by the temperature $\beta$ appearing in the driving term. These have simple expressions \cite{NESS,NESS2}, \begin{equation}
\begin{split}
    \braket{J_{E,R}^n}_c =\frac{n!c\pi}{12\beta^{n+1}},\hspace{0.5cm} 
    \braket{J_{E,L}^n}_c =(-1)^n\frac{n!c\pi}{12\beta^{n+1}}
    \end{split}
    \label{JJ}
\end{equation}
where the inverse temperature appearing in both is the same because we are considering a homogeneous setup. Plugging Eq. \eqref{JJ} into \eqref{eq:ehcft}, we have
\begin{equation}
    \braket{K_A^n}^{\rho_A}_c = 2t\frac{c\pi n!}{6\beta}.
\end{equation}
For $n=2$, it reproduces the same linear growth of the entanglement entropy $S_A$ predicted in Ref. \cite{quench2}. 
The equivalence between the quench dynamics of $S_A(t)$ and the second cumulant of the entanglement Hamiltonian, $\langle K_A^2\rangle_c$ (the so-called \emph{capacity of entanglement}), was already pointed out in Refs.~\cite{deboer2019,Arias2023}, providing an independent confirmation of our results.
It is worth noting that the connection between the entanglement Hamiltonian and current operators in the conformal setting is already implicit in the general solution of Ref.~\cite{Cardy_2016}, although, to the best of our knowledge, the explicit form~\eqref{eq:ehcft} has not previously appeared in the literature. Our main interest here, however, is not the conformal result itself, but rather its generality: Eq.~\eqref{eq:ehcft} emerges naturally as the conformal limit of a framework that applies to generic integrable models.

\section{Conclusions}\label{sec:concl}
We have employed the framework of space–time duality to investigate the dynamics of the cumulants of the entanglement Hamiltonian following a quantum quench in integrable systems. Our main result is that, in the ballistic regime, these cumulants admit a remarkably simple interpretation in terms of the cumulants of suitably defined current operators. This establishes a new and direct connection between entanglement dynamics and transport, providing a thermodynamic counterpart to the quasiparticle picture.

This perspective sheds new light on the origin of the different behavior observed in free and interacting integrable models. In particular, it naturally explains why the quasiparticle picture successfully describes the von Neumann entropy but fails for Rényi entropies in interacting systems, while at the same time clarifying why it remains exact in free theories and, more generally, in conformal field theories where right- and left-moving excitations decouple. Our results therefore suggest that the transport of conserved currents could provide the more fundamental mechanism underlying entanglement growth in interacting integrable systems.

An important open problem is to move beyond the description of the cumulants and obtain a genuine operator-level characterization of the entanglement Hamiltonian in interacting models. Such a formulation would provide a much deeper understanding of the microscopic origin of entanglement dynamics and could establish a direct bridge between the thermodynamic description developed here and the operator-based quasiparticle picture available in free systems. More generally, it would be interesting to explore whether similar ideas can be extended beyond integrability, thereby providing a broader framework for understanding the interplay between transport and entanglement in generic quantum many-body systems.
Finally, it would also be interesting to investigate whether some of our results can be generalised to the (cumulants of the) negativity Hamiltonian \cite{murciano2022negativity,rottoli2023finite,travaglino2025}, thereby providing a relation between  mixed-state entanglement and transport.

\paragraph{Acknowledgements ---}
We thank Katja Klobas, Bruno Bertini, David Horvath and Fabian Essler for interesting discussions, and Colin Rylands for collaborations on related topics. Both authors acknowledge support by the ERC-AdG grant MOSE No. 101199196. 

\bibliographystyle{ytphys}
\bibliography{bibliography}

@article{calabrese2016introduction,
  title={Introduction to ‘{Quantum} integrability in out of equilibrium systems’},
  author={Calabrese, Pasquale and Essler, Fabian H L and Mussardo, Giuseppe},
  journal={J. Stat. Mech.},
  number={6},
  pages={064001},
  year={2016},
  publisher={{IOP}},
  doi={10.1088/1742-5468/2016/06/064001}
}

@article{carollo2022,
  title = {Entangled multiplets and spreading of quantum correlations in a continuously monitored tight-binding chain},
  author = {Carollo, Federico and Alba, Vincenzo},
  journal = {Phys. Rev. B},
  volume = {106},
  issue = {22},
  pages = {L220304},
  numpages = {7},
  year = {2022},
  
  publisher = {American Physical Society},
  doi = {10.1103/PhysRevB.106.L220304},
  url = {https://link.aps.org/doi/10.1103/PhysRevB.106.L220304}
}

@article{bertini2025exactly,
  title = {Exactly solvable quantum many-body dynamics from space-time duality},
  author = {Bertini, Bruno and Claeys, Pieter W. and Prosen, Toma\ifmmode \check{z}\else \v{z}\fi{}},
  journal = {Rev. Mod. Phys.},
  volume = {98},
  issue = {2},
  pages = {025001},
  numpages = {63},
  year = {2026},
  publisher = {American Physical Society},
  doi = {10.1103/yx73-dk86},
  url = {https://link.aps.org/doi/10.1103/yx73-dk86}
}

@article{travaglino2026spacetimeduality,
      title={Space-time duality approach to (inhomogeneous) integrable quenches}, 
      author={Riccardo Travaglino and Pasquale Calabrese and Katja Klobas and Bruno Bertini},
      year={2026},
      eprint={2606.20445},
      archivePrefix={arXiv},
}

@article{Bonsignori2026,
author={Bonsignori, Riccarda
and Eisler, Viktor},
title={Entanglement {Hamiltonian} after a local quench},
journal={J. High Energy Phys.},
year={2026},
day={23},
volume= {02},
pages={232},
issn={1029-8479},
doi={10.1007/JHEP02(2026)232},
url={https://doi.org/10.1007/JHEP02(2026)232}
}

@article{Eisler_2025,
doi = {10.1088/1742-5468/ad9c4f},
url = {https://doi.org/10.1088/1742-5468/ad9c4f},
year = {2025},

publisher = {IOP Publishing},
number = {1},
pages = {013101},
author = {Eisler, Viktor},
title = {On the {Bisognano–Wichmann entanglement Hamiltonian} of nonrelativistic fermions},
journal = {J. Stat. Mech.},
}

@article{BERNARD2025117185,
title = {Entanglement {Hamiltonian} and orthogonal polynomials},
journal = {Nucl. Phys. B},
volume = {1020},
pages = {117185},
year = {2025},
issn = {0550-3213},
doi = {https://doi.org/10.1016/j.nuclphysb.2025.117185},
url = {https://www.sciencedirect.com/science/article/pii/S0550321325003943},
author = {Pierre-Antoine Bernard and Riccarda Bonsignori and Viktor Eisler and Gilles Parez and Luc Vinet},
}

@article{eisler2024,
  title = {Entanglement {Hamiltonian} of a nonrelativistic Fermi gas},
  author = {Eisler, Viktor},
  journal = {Phys. Rev. B},
  volume = {109},
  issue = {20},
  pages = {L201113},
  numpages = {6},
  year = {2024},
  publisher = {American Physical Society},
  doi = {10.1103/PhysRevB.109.L201113},
  url = {https://link.aps.org/doi/10.1103/PhysRevB.109.L201113}
}

@article{Bonsignori_2024,
doi = {10.1088/1751-8121/ad5501},
url = {https://doi.org/10.1088/1751-8121/ad5501},
year = {2024},

publisher = {IOP Publishing},
volume = {57},
number = {27},
pages = {275001},
author = {Bonsignori, Riccarda and Eisler, Viktor},
title = {Entanglement {Hamiltonian} for inhomogeneous free fermions},
journal = {J. Phys. A},
}

@article{toldin2018,
  title = {Entanglement {Hamiltonian} of Interacting Fermionic Models},
  author = {Parisen Toldin, Francesco and Assaad, Fakher F.},
  journal = {Phys. Rev. Lett.},
  volume = {121},
  issue = {20},
  pages = {200602},
  numpages = {6},
  year = {2018},
  
  publisher = {American Physical Society},
  doi = {10.1103/PhysRevLett.121.200602},
  url = {https://link.aps.org/doi/10.1103/PhysRevLett.121.200602}
}

@article{Caux_2011,
doi = {10.1088/1742-5468/2011/02/P02023},
url = {https://doi.org/10.1088/1742-5468/2011/02/P02023},
year = {2011},

publisher = {},
number = {02},
pages = {P02023},
author = {Caux, Jean-Sébastien and Mossel, Jorn},
title = {Remarks on the notion of quantum integrability},
journal = {J. Stat. Mech.},
}

@article{bertini2022entanglement,
  title = {Entanglement Negativity and Mutual Information after a Quantum Quench: Exact Link from Space-Time Duality},
  author = {Bertini, Bruno and Klobas, Katja and Lu, Tsung-Cheng},
  journal = {Phys. Rev. Lett.},
  volume = {129},
  issue = {14},
  pages = {140503},
  numpages = {8},
  year = {2022},
  
  publisher = {American Physical Society},
  doi = {10.1103/PhysRevLett.129.140503},
  url = {https://link.aps.org/doi/10.1103/PhysRevLett.129.140503}
}

@article{carollo_alba2022,
  title = {Dissipative quasiparticle picture for quadratic Markovian open quantum systems},
  author = {Carollo, Federico and Alba, Vincenzo},
  journal = {Phys. Rev. B},
  volume = {105},
  issue = {14},
  pages = {144305},
  numpages = {12},
  year = {2022},
  
  publisher = {American Physical Society},
  doi = {10.1103/PhysRevB.105.144305},
  url = {https://link.aps.org/doi/10.1103/PhysRevB.105.144305}
}

@article{bertini2024dynamics,
  title = {Dynamics of charge fluctuations from asymmetric initial states},
  author = {Bertini, Bruno and Klobas, Katja and Collura, Mario and Calabrese, Pasquale and Rylands, Colin},
  journal = {Phys. Rev. B},
  volume = {109},
  issue = {18},
  pages = {184312},
  numpages = {28},
  year = {2024},
  
  publisher = {American Physical Society},
  doi = {10.1103/PhysRevB.109.184312},
  url = {https://link.aps.org/doi/10.1103/PhysRevB.109.184312}
}

@article{rottoli2024entanglementhamiltoniansquasiparticlepicture,
  title = {Entanglement {Hamiltonians} and the quasiparticle picture},
  author = {Rottoli, Federico and Rylands, Colin and Calabrese, Pasquale},
  journal = {Phys. Rev. B},
  volume = {111},
  issue = {14},
  pages = {L140302},
  numpages = {7},
  year = {2025},
  
  publisher = {American Physical Society},
  doi = {10.1103/PhysRevB.111.L140302},
  url = {https://link.aps.org/doi/10.1103/PhysRevB.111.L140302}
}

@article{travaglino2024,
doi = {10.1088/1742-5468/adb7d3},
url = {https://dx.doi.org/10.1088/1742-5468/adb7d3},
year = {2025},

publisher = {IOP Publishing},
number = {3},
pages = {033102},
author = {Travaglino, Riccardo and Rylands, Colin and Calabrese, Pasquale},
title = {{Quasiparticle picture for entanglement hamiltonians in higher dimensions}},
journal = {J. Stat. Mech.},
}

@article{Travaglino2025measurements,
doi = {10.1088/1742-5468/ae09a0},
url = {https://doi.org/10.1088/1742-5468/ae09a0},
year = {2025},
publisher = {IOP Publishing},
number = {12},
pages = {123101},
author = {Travaglino, Riccardo and Rylands, Colin and Calabrese, Pasquale},
title = {{Quench dynamics of entanglement entropy under projective charge measurements: the free fermion case}},
journal = {J. Stat. Mech.}
}

@article{travaglino2026dissipative,
doi = {10.1088/1742-5468/ae6c14},
url = {https://doi.org/10.1088/1742-5468/ae6c14},
year = {2026},
publisher = {IOP Publishing},
number = {6},
pages = {063103},
author = {Travaglino, Riccardo and Rottoli, Federico and Calabrese, Pasquale},
title = {{Entanglement Hamiltonians in dissipative free fermions and the time-dependent GGE}},
journal = {J. Stat. Mech.},
}

@article{travaglino2025,
doi = {10.1088/1742-5468/adfe58},
url = {https://doi.org/10.1088/1742-5468/adfe58},
year = {2025},

publisher = {IOP Publishing},
number = {9},
pages = {093103},
author = {Travaglino, Riccardo and Rylands, Colin and Calabrese, Pasquale},
title = {{Quench dynamics of negativity Hamiltonians}},
journal = {J. Stat. Mech.},
}

@article{Caux_2016,
doi = {10.1088/1742-5468/2016/06/064006},
url = {https://doi.org/10.1088/1742-5468/2016/06/064006},
year = {2016},
publisher = {IOP Publishing and SISSA},
number = {6},
pages = {064006},
author = {Caux, Jean-Sébastien},
title = {The Quench Action},
journal = {J. Stat. Mech.},
}

@article{Rottoli_2024,
url = {https://doi.org/10.1088/1742-5468/ad4860},
year = {2024},
number = {6},
pages = {063102},
author = {Rottoli, Federico and Fossati, Michele and Calabrese, Pasquale},
title = {Entanglement {Hamiltonian} in the non-{Hermitian SSH} model},
journal = {J. Stat. Mech.}
}

@article{Rottoli_2022,
   title={Entanglement {Hamiltonian during a domain wall melting in the free Fermi chain}},
   ISSN={1742-5468},
   url={http://dx.doi.org/10.1088/1742-5468/ac72a1},
   DOI={10.1088/1742-5468/ac72a1},
   number={6},
   journal={J. Stat. Mech.},
   publisher={IOP Publishing},
   author={Rottoli, Federico and Scopa, Stefano and Calabrese, Pasquale},
   year={2022},
   pages={063103} }

@article{Dubail_2017,
doi = {10.1088/1751-8121/aa6f38},
url = {https://doi.org/10.1088/1751-8121/aa6f38},
year = {2017},
publisher = {IOP Publishing},
volume = {50},
number = {23},
pages = {234001},
author = {Dubail, J},
title = {Entanglement scaling of operators: a conformal field theory approach, with a glimpse of simulability of long-time dynamics in 1+1d},
journal = {J. Phys. A}
}

@article{caux2013,
  title = {Time Evolution of Local Observables After Quenching to an Integrable Model},
  author = {Caux, Jean-S\'ebastien and Essler, Fabian H. L.},
  journal = {Phys. Rev. Lett.},
  volume = {110},
  issue = {25},
  pages = {257203},
  numpages = {5},
  year = {2013},
  
  publisher = {American Physical Society},
  doi = {10.1103/PhysRevLett.110.257203},
  url = {https://link.aps.org/doi/10.1103/PhysRevLett.110.257203}
}

@article{bertini2018entanglement,
doi = {10.1088/1751-8121/aad82e},
url = {https://dx.doi.org/10.1088/1751-8121/aad82e},
year = {2018},

publisher = {IOP Publishing},
volume = {51},
number = {39},
pages = {39LT01},
author = {Bertini, Bruno and Fagotti, Maurizio and Piroli, Lorenzo and Calabrese, Pasquale},
title = {Entanglement evolution and generalised hydrodynamics: noninteracting systems},
journal = {J. Phys. A},
}

@article{alba_quench_action_2017,
  title = {Quench action and {R}\'enyi entropies in integrable systems},
  author = {Alba, Vincenzo and Calabrese, Pasquale},
  journal = {Phys. Rev. B},
  volume = {96},
  issue = {11},
  pages = {115421},
  numpages = {8},
  year = {2017},
  
  publisher = {American Physical Society},
  doi = {10.1103/PhysRevB.96.115421},
  url = {https://link.aps.org/doi/10.1103/PhysRevB.96.115421}
}

@article{bertini2022growth,
  title = {Growth of {R}\'enyi Entropies in Interacting Integrable Models and the Breakdown of the Quasiparticle Picture},
  author = {Bertini, Bruno and Klobas, Katja and Alba, Vincenzo and Lagnese, Gianluca and Calabrese, Pasquale},
  journal = {Phys. Rev. X},
  volume = {12},
  issue = {3},
  pages = {031016},
  numpages = {20},
  year = {2022},
  publisher = {American Physical Society},
  doi = {10.1103/PhysRevX.12.031016},
  url = {https://link.aps.org/doi/10.1103/PhysRevX.12.031016}
}

@article{yangyang,
    author = {Yang, C. N. and Yang, C. P.},
    title = {Thermodynamics of a One‐Dimensional System of Bosons with Repulsive Delta‐Function Interaction},
    journal = {J. Math. Phys.},
    volume = {10},
    number = {7},
    pages = {1115},
    year = {1969},
  
    issn = {0022-2488},
    doi = {10.1063/1.1664947},
    url = {https://doi.org/10.1063/1.1664947},
    }

@article{Essler_2016,
doi = {10.1088/1742-5468/2016/06/064002},
url = {https://dx.doi.org/10.1088/1742-5468/2016/06/064002},
year = {2016},

publisher = {IOP Publishing and SISSA},
number = {6},
pages = {064002},
author = {Essler, Fabian H L and Fagotti, Maurizio},
title = {Quench dynamics and relaxation in isolated integrable quantum spin chains},
journal = {J. Stat. Mech.},

}

@article{dipasquale2026,
      title={Entanglement dynamics after quenches with inhomogeneous Hamiltonians}, 
      author={Andrea Di Pasquale and Federico Rottoli and Vincenzo Alba},
      year={2026},
      eprint={2605.01595},
      archivePrefix={arXiv},
      primaryClass={cond-mat.stat-mech},
}

@article{Bastianello_2020,
   title={Entanglement spreading and quasiparticle picture beyond the pair structure},
   volume={8},
   ISSN={2542-4653},
   url={http://dx.doi.org/10.21468/SciPostPhys.8.3.045},
   DOI={10.21468/scipostphys.8.3.045},
   number={3},
   journal={SciPost Phys.},
   publisher={SciPost},
   author={Bastianello, Alvise and Collura, Mario},
   year={2020} }

@article{santalla2023,
  title = {Entanglement links and the quasiparticle picture},
  author = {Santalla, Silvia N. and Ram\'{\i}rez, Giovanni and Roy, Sudipto Singha and Sierra, Germ\'an and Rodr\'{\i}guez-Laguna, Javier},
  journal = {Phys. Rev. B},
  volume = {107},
  issue = {12},
  pages = {L121114},
  numpages = {6},
  year = {2023},
  publisher = {American Physical Society},
  doi = {10.1103/PhysRevB.107.L121114},
  url = {https://link.aps.org/doi/10.1103/PhysRevB.107.L121114}
}

@article{Casini_2011,
   title={Towards a derivation of holographic entanglement entropy},
   ISSN={1029-8479},
   url={http://dx.doi.org/10.1007/JHEP05(2011)036},
   DOI={10.1007/jhep05(2011)036},
   volume={05},
   journal={J. High Energy Phys.},
   publisher={Springer Science and Business Media LLC},
   author={Casini, Horacio and Huerta, Marina and Myers, Robert C.},
   pages={036},
   year={2011}}

@article{fulgado2025,
      title={A quasi-particle picture for entanglement cones and horizons in analogue cosmology}, 
      author={Carlos Fulgado-Claudio and Alejandro Bermudez},
      year={2025},
      eprint={2503.19183},
      archivePrefix={arXiv},
      primaryClass={quant-ph},
}

@article{Dalmonte_2018,
   title={{Quantum simulation and spectroscopy of entanglement Hamiltonians}},
   volume={14},
   ISSN={1745-2481},
   url={http://dx.doi.org/10.1038/s41567-018-0151-7},
   DOI={10.1038/s41567-018-0151-7},
   number={8},
   journal={Nature Physics},
   publisher={Springer Science and Business Media LLC},
   author={Dalmonte, M. and Vermersch, B. and Zoller, P.},
   year={2018},
 pages={827} }

@article{Kokail_2021,
   title={{Quantum variational learning of the entanglement Hamiltonian}},
   volume={127},
   ISSN={1079-7114},
   url={http://dx.doi.org/10.1103/PhysRevLett.127.170501},
   DOI={10.1103/physrevlett.127.170501},
   number={17},
   pages={170501},
   journal={Phys. Rev. Lett.},
   publisher={American Physical Society (APS)},
   author={Kokail, Christian and Sundar, Bhuvanesh and Zache, Torsten V. and Elben, Andreas and Vermersch, Benoît and Dalmonte, Marcello and van Bijnen, Rick and Zoller, Peter},
   year={2021} }

@article{yang_2026,
  title = {{Limits of the lattice Bisognano-Wichmann form for entanglement Hamiltonians: A quantum Monte Carlo study}},
  author = {Yang, Siyi and Ding, Yi-Ming and Yan, Zheng},
  journal = {Phys. Rev. B},
  volume = {113},
  issue = {23},
  pages = {235115},
  numpages = {22},
  year = {2026},
  publisher = {American Physical Society},
  doi = {10.1103/5tk7-dxqk},
  url = {https://link.aps.org/doi/10.1103/5tk7-dxqk}
}

@article{klich-2018,
  title = {Entanglement Hamiltonians and entropy in (1+1)-dimensional chiral fermion systems},
  author = {Klich, Israel and Vaman, Diana and Wong, Gabriel},
  journal = {Phys. Rev. B},
  volume = {98},
  issue = {3},
  pages = {035134},
  numpages = {12},
  year = {2018},
  publisher = {American Physical Society},
  doi = {10.1103/PhysRevB.98.035134},
  url = {https://link.aps.org/doi/10.1103/PhysRevB.98.035134}
}

@article{horodecki2009,
  title = {Quantum entanglement},
  author = {Horodecki, Ryszard and Horodecki, Pawe\l{} and Horodecki, Micha\l{} and Horodecki, Karol},
  journal = {Rev. Mod. Phys.},
  volume = {81},
  issue = {2},
  pages = {865},
  numpages = {0},
  year = {2009},
  
  publisher = {American Physical Society},
  doi = {10.1103/RevModPhys.81.865},
  url = {https://link.aps.org/doi/10.1103/RevModPhys.81.865}
}

@article{damico2008,
  title = {Entanglement in many-body systems},
  author = {Amico, Luigi and Fazio, Rosario and Osterloh, Andreas and Vedral, Vlatko},
  journal = {Rev. Mod. Phys.},
  volume = {80},
  issue = {2},
  pages = {517},
  numpages = {0},
  year = {2008},
  
  publisher = {American Physical Society},
  doi = {10.1103/RevModPhys.80.517},
  url = {https://link.aps.org/doi/10.1103/RevModPhys.80.517}
}

@article{fagotti2008evolution,
  title = {{Evolution of entanglement entropy following a quantum quench: Analytic results for the $XY$ chain in a transverse magnetic field}},
  author = {Fagotti, Maurizio and Calabrese, Pasquale},
  journal = {Phys. Rev. A},
  volume = {78},
  issue = {1},
  pages = {010306},
  numpages = {4},
  year = {2008},
  publisher = {American Physical Society},
  doi = {10.1103/PhysRevA.78.010306},
  url = {https://link.aps.org/doi/10.1103/PhysRevA.78.010306}
}

@article{turkeshi2023,
  title = {Entanglement and correlation spreading in non-Hermitian spin chains},
  author = {Turkeshi, Xhek and Schir\'o, Marco},
  journal = {Phys. Rev. B},
  volume = {107},
  issue = {2},
  pages = {L020403},
  numpages = {6},
  year = {2023},
  
  publisher = {American Physical Society},
  doi = {10.1103/PhysRevB.107.L020403},
  url = {https://link.aps.org/doi/10.1103/PhysRevB.107.L020403}
}

@article{Turkeshi2022,
  title = {Enhanced entanglement negativity in boundary-driven monitored fermionic chains},
  author = {Turkeshi, Xhek and Piroli, Lorenzo and Schir\'o, Marco},
  journal = {Phys. Rev. B},
  volume = {106},
  issue = {2},
  pages = {024304},
  numpages = {11},
  year = {2022},
  
  publisher = {American Physical Society},
  doi = {10.1103/PhysRevB.106.024304},
  url = {https://link.aps.org/doi/10.1103/PhysRevB.106.024304}
}

@Article{klobas2021entanglement,
	title={{Entanglement dynamics in Rule 54: Exact results and quasiparticle picture}},
	author={Katja Klobas and Bruno Bertini},
	journal={SciPost Phys.},
	volume={11},
	pages={107},
	year={2021},
	publisher={SciPost},
	doi={10.21468/SciPostPhys.11.6.107},
	url={https://scipost.org/10.21468/SciPostPhys.11.6.107},
}

@article{Murciano2022,
   title={Quench Dynamics of Rényi Negativities and the Quasiparticle Picture},
   ISBN={9783031039980},
   ISSN={2364-9062},
   url={http://dx.doi.org/10.1007/978-3-031-03998-0_14},
   DOI={10.1007/978-3-031-03998-0_14},
   journal ={Entanglement in Spin Chains},
   publisher={Springer International Publishing},
   author={Murciano, Sara and Alba, Vincenzo and Calabrese, Pasquale},
   year={2022},
   pages={397} }

@Article{alba2019entanglement,
	title={{Entanglement evolution and generalised hydrodynamics: interacting integrable systems}},
	author={Vincenzo Alba and Bruno Bertini and Maurizio Fagotti},
	journal={SciPost Phys.},
	volume={7},
	pages={005},
	year={2019},
	publisher={SciPost},
	doi={10.21468/SciPostPhys.7.1.005},
	url={https://scipost.org/10.21468/SciPostPhys.7.1.005},
}

@Article{doyon2020lecture,
	title={{Lecture notes on Generalised Hydrodynamics}},
	author={Benjamin Doyon},
	journal={SciPost Phys. Lect. Notes},
	pages={18},
	year={2020},
	publisher={SciPost},
	doi={10.21468/SciPostPhysLectNotes.18},
	url={https://scipost.org/10.21468/SciPostPhysLectNotes.18},
}

@article{EH,
author = {Dalmonte, Marcello and Eisler, Viktor and Falconi, Marco and Vermersch, Benoît},
title = {{Entanglement Hamiltonians: From Field Theory to Lattice Models and Experiments}},
journal = {Ann. Phys.},
volume = {534},
number = {11},
pages = {2200064},
doi = {https://doi.org/10.1002/andp.202200064},
year= {2022}

}

@article{arealaws,
  title = {Colloquium: Area laws for the entanglement entropy},
  author = {Eisert, J. and Cramer, M. and Plenio, M. B.},
  journal = {Rev. Mod. Phys.},
  volume = {82},
  issue = {1},
  pages = {277--306},
  numpages = {0},
  year = {2010},
  
  publisher = {American Physical Society},
  doi = {10.1103/RevModPhys.82.277},
  url = {https://link.aps.org/doi/10.1103/RevModPhys.82.277}
}

@article{
alba1,
author = {Vincenzo Alba  and Pasquale Calabrese },
title = {Entanglement and thermodynamics after a quantum quench in integrable systems},
journal = {PNAS},
volume = {114},
number = {30},
pages = {7947},
year = {2017},
doi = {10.1073/pnas.1703516114}
}

@article{Eisler_2019,
   title={{On the continuum limit of the entanglement Hamiltonian}},
   ISSN={1742-5468},
   url={http://dx.doi.org/10.1088/1742-5468/ab1f0e},
   DOI={10.1088/1742-5468/ab1f0e},
   number={7},
   journal={J. Stat. Mech.},
   publisher={IOP Publishing},
   author={Eisler, Viktor and Tonni, Erik and Peschel, Ingo},
   year={2019},
 pages={073101} }

@article{Di_Giulio_2020,
   title={{On entanglement Hamiltonians of an interval in massless harmonic chains}},
   ISSN={1742-5468},
   url={http://dx.doi.org/10.1088/1742-5468/ab7129},
   DOI={10.1088/1742-5468/ab7129},
   number={3},
   journal={J. Stat. Mech.},
   publisher={IOP Publishing},
   author={Di Giulio, Giuseppe and Tonni, Erik},
   year={2020},
 pages={033102} }

@article{Mintchev_2021,
   title={{Modular Hamiltonians for the massless Dirac field in the presence of a boundary}},
   volume={03},
   ISSN={1029-8479},
   url={http://dx.doi.org/10.1007/JHEP03(2021)204},
   DOI={10.1007/jhep03(2021)204},
   number={3},
   journal={J. High Energy Phys.},
   publisher={Springer Science and Business Media LLC},
   author={Mintchev, Mihail and Tonni, Erik},
   year={2021},
   pages={204} }

@article{Li_2024,
   title={{Numerical investigations of the extensive entanglement Hamiltonian in quantum spin ladders}},
   volume={3},
   ISSN={2731-6106},
   url={http://dx.doi.org/10.1007/s44214-024-00056-2},
   DOI={10.1007/s44214-024-00056-2},
   number={1},
   journal={Quantum Front.},
   publisher={Springer Science and Business Media LLC},
   author={Li, Chengshu and Li, Xingyu and Zhou, Yi-Neng},
   pages={9},
   year={2024} }

@Article{alba2,
	title={{Entanglement dynamics after quantum quenches in generic integrable systems}},
	author={Vincenzo Alba and Pasquale Calabrese},
	journal={SciPost Phys.},
	volume={4},
	pages={017},
	year={2018},
	publisher={SciPost},
	doi={10.21468/SciPostPhys.4.3.017},
	url={https://scipost.org/10.21468/SciPostPhys.4.3.017},
}

@book{takahashi1999book,
    author = "Takahashi, M.",
    title = "{Thermodynamics of One-Dimensional Solvable Models}",
    doi = "10.1017/cbo9780511524332",
    isbn = "978-0-511-52433-2",
    publisher = "Cambridge University Press",
    year = "1999"
}

@article{Yang,
  title = "{Some Exact Results for the Many-Body Problem in one Dimension with Repulsive Delta-Function Interaction}",
  author = {Yang, C. N.},
  journal = {Phys. Rev. Lett.},
  volume = {19},
  issue = {23},
  pages = {1312},
  numpages = {0},
  year = {1967},
  
  publisher = {American Physical Society},
  doi = {10.1103/PhysRevLett.19.1312},
  url = {https://link.aps.org/doi/10.1103/PhysRevLett.19.1312}
}

@book{Mussardo:2010mgq,
    author = "Mussardo, Giuseppe",
    title = "{Statistical field theory}: {an introduction to exactly solved models in statistical physics}",
    publisher = "Oxford Univ. Press",
    address = "New York, NY",
    year = "2010"
}

@article{Zamolodchikov1989,
    author = "Zamolodchikov, A. B.",
    title = "{Thermodynamic Bethe Ansatz in Relativistic Models. Scaling Three State Potts and Lee-yang Models}",
    reportNumber = "ITEP-89-144",
    doi = "10.1016/0550-3213(90)90333-9",
    journal = "Nucl. Phys. B",
    volume = "342",
    pages = "695",
    year = "1990"
}

@article{castroalvaredo2016emergent,
  title = {Emergent Hydrodynamics in Integrable Quantum Systems Out of Equilibrium},
  author = {Castro-Alvaredo, Olalla A. and Doyon, Benjamin and Yoshimura, Takato},
  journal = {Phys. Rev. X},
  volume = {6},
  issue = {4},
  pages = {041065},
  numpages = {17},
  year = {2016},
  publisher = {American Physical Society},
  doi = {10.1103/PhysRevX.6.041065},
  url = {https://link.aps.org/doi/10.1103/PhysRevX.6.041065}
}

@article{RevModPhys.83.863,
  title = {Colloquium: Nonequilibrium dynamics of closed interacting quantum systems},
  author = {Polkovnikov, Anatoli and Sengupta, Krishnendu and Silva, Alessandro and Vengalattore, Mukund},
  journal = {Rev. Mod. Phys.},
  volume = {83},
  issue = {3},
  pages = {863},
  numpages = {0},
  year = {2011},
  publisher = {American Physical Society},
  doi = {10.1103/RevModPhys.83.863},
  url = {https://link.aps.org/doi/10.1103/RevModPhys.83.863}
}

@article{Calabrese_2009,
doi = {10.1088/1751-8113/42/50/504005},
url = {https://dx.doi.org/10.1088/1751-8113/42/50/504005},
year = {2009},

publisher = {},
volume = {42},
number = {50},
pages = {504005},
author = {Pasquale Calabrese and John Cardy},
title = {Entanglement entropy and conformal field theory},
journal = {J. Phys. A},

}

@article{quench1,
  title = {Time Dependence of Correlation Functions Following a Quantum Quench},
  author = {Calabrese, Pasquale and Cardy, John},
  journal = {Phys. Rev. Lett.},
  volume = {96},
  issue = {13},
  pages = {136801},
  numpages = {4},
  year = {2006},
  
  publisher = {American Physical Society},
  doi = {10.1103/PhysRevLett.96.136801},
  url = {https://link.aps.org/doi/10.1103/PhysRevLett.96.136801}
}

@article{quench2,
doi = {10.1088/1742-5468/2005/04/P04010},
url = {https://dx.doi.org/10.1088/1742-5468/2005/04/P04010},
year = {2005},
publisher = {},
number = {04},
pages = {P04010},
author = {Pasquale Calabrese and John Cardy},
title = {Evolution of entanglement entropy in one-dimensional systems},
journal = {J. Stat. Mech.},

}

@article{Calabrese_2009_1,
doi = {10.1088/1751-8121/42/50/500301},
url = {https://dx.doi.org/10.1088/1751-8121/42/50/500301},
year = {2009},
publisher = {},
volume = {42},
number = {50},
pages = {500301},
author = {Pasquale Calabrese and John Cardy and Benjamin Doyon},
title = {Entanglement entropy in extended quantum systems},
journal = {J. Phys. A}
}

@article{bertini2016transport,
  title = {{Transport in Out-of-Equilibrium $XXZ$ Chains: Exact Profiles of Charges and Currents}},
  author = {Bertini, Bruno and Collura, Mario and De Nardis, Jacopo and Fagotti, Maurizio},
  journal = {Phys. Rev. Lett.},
  volume = {117},
  issue = {20},
  pages = {207201},
  numpages = {8},
  year = {2016},

  publisher = {American Physical Society},
  doi = {10.1103/PhysRevLett.117.207201},
  url = {https://link.aps.org/doi/10.1103/PhysRevLett.117.207201}
}

@article{Bisognano:1975ih,
    author = "Bisognano, J. J and Wichmann, E. H.",
    title = "{On the Duality Condition for a Hermitian Scalar Field}",
    doi = "10.1063/1.522605",
    journal = "J. Math. Phys.",
    volume = "16",
    pages = "985",
    year = "1975"
}

@article{Bisognano:1976za,
    author = "Bisognano, J. J and Wichmann, E. H.",
    title = "{On the Duality Condition for Quantum Fields}",
    doi = "10.1063/1.522898",
    journal = "J. Math. Phys.",
    volume = "17",
    pages = "303",
    year = "1976"
}

@article{ehFF1,
doi = {10.1088/1751-8121/aa76b5},
url = {https://dx.doi.org/10.1088/1751-8121/aa76b5},
year = {2017},
publisher = {IOP Publishing},
volume = {50},
number = {28},
pages = {284003},
author = {Viktor Eisler and Ingo Peschel},
title = {Analytical results for the entanglement {Hamiltonian} of a free-fermion chain},
journal = {J. Phys. A},

}

@article{EHFF2,
doi = {10.1088/1742-5468/aace2b},
url = {https://dx.doi.org/10.1088/1742-5468/aace2b},
year = {2018},

publisher = {IOP Publishing and SISSA},
number = {10},
pages = {104001},
author = {Viktor Eisler and Ingo Peschel},
title = {Properties of the entanglement {Hamiltonian} for finite free-fermion chains},
journal = {J. Stat. Mech.},

}

@article{DAlessio:2015qtq,
    author = "D'Alessio, Luca and Kafri, Yariv and Polkovnikov, Anatoli and Rigol, Marcos",
    title = "{From quantum chaos and eigenstate thermalization to statistical mechanics and thermodynamics}",
    doi = "10.1080/00018732.2016.1198134",
    journal = "Adv. Phys.",
    volume = "65",
    number = "3",
    pages = "239",
    year = "2016"
}

@article{Gogolin_2016,
doi = {10.1088/0034-4885/79/5/056001},
url = {https://dx.doi.org/10.1088/0034-4885/79/5/056001},
year = {2016},
publisher = {IOP Publishing},
volume = {79},
number = {5},
pages = {056001},
author = {Christian Gogolin and Jens Eisert},
title = {Equilibration, thermalisation, and the emergence of statistical mechanics in closed quantum systems},
journal = {Rep. Prog. Phys.},
}

@article{Cardy_2016,
doi = {10.1088/1742-5468/2016/12/123103},
url = {https://dx.doi.org/10.1088/1742-5468/2016/12/123103},
year = {2016},
publisher = {IOP Publishing and SISSA},
number = {12},
pages = {123103},
author = {John Cardy and Erik Tonni},
title = {{Entanglement Hamiltonians in two-dimensional conformal field theory}},
journal = {J. Stat. Mech.},

}

@article{digiulio2019entanglement,
   title="{Entanglement Hamiltonians in 1D free lattice models after a global quantum quench}",
   ISSN={1742-5468},
   url={http://dx.doi.org/10.1088/1742-5468/ab4e8f},
   DOI={10.1088/1742-5468/ab4e8f},
   number={12},
   journal={J. Stat. Mech.},
   publisher={IOP Publishing},
   author={Di Giulio, Giuseppe and Arias, Raúl and Tonni, Erik},
   year={2019}, pages={123103} }

@article{zhu2020entanglement,
  title = "{Entanglement Hamiltonian of many-body dynamics in strongly correlated systems}",
  author = {Zhu, W. and Huang, Zhoushen and He, Yin-Chen and Wen, Xueda},
  journal = {Phys. Rev. Lett.},
  volume = {124},
  issue = {10},
  pages = {100605},
  numpages = {6},
  year = {2020},
  publisher = {American Physical Society},
  doi = {10.1103/PhysRevLett.124.100605},
  url = {https://link.aps.org/doi/10.1103/PhysRevLett.124.100605}
}

@article{coser2014entanglement,
   title={Entanglement negativity after a global quantum quench},
   ISSN={1742-5468},
   url={http://dx.doi.org/10.1088/1742-5468/2014/12/P12017},
   DOI={10.1088/1742-5468/2014/12/p12017},
   number={12},
   journal={J. Stat. Mech.},
   publisher={IOP Publishing},
   author={Coser, Andrea and Tonni, Erik and Calabrese, Pasquale},
   year={2014},
 pages={P12017} }

@article{alba2019quantum,
   title={Quantum information dynamics in multipartite integrable systems},
   volume={126},
   ISSN={1286-4854},
   url={http://dx.doi.org/10.1209/0295-5075/126/60001},
   DOI={10.1209/0295-5075/126/60001},
   number={6},
   journal={Europhys. Lett.},
   publisher={IOP Publishing},
   author={Alba, Vincenzo and Calabrese, Pasquale},
   year={2019}, pages={60001} }

@article{bertini2023nonequilibrium,
  title = {Nonequilibrium Full Counting Statistics and Symmetry-Resolved Entanglement from Space-Time Duality},
  author = {Bertini, Bruno and Calabrese, Pasquale and Collura, Mario and Klobas, Katja and Rylands, Colin},
  journal = {Phys. Rev. Lett.},
  volume = {131},
  issue = {14},
  pages = {140401},
  numpages = {7},
  year = {2023},
  publisher = {American Physical Society},
  doi = {10.1103/PhysRevLett.131.140401},
  url = {https://link.aps.org/doi/10.1103/PhysRevLett.131.140401}
}

@article{parez2021quasiparticle,
  title = {Quasiparticle dynamics of symmetry-resolved entanglement after a quench: Examples of conformal field theories and free fermions},
  author = {Parez, Gilles and Bonsignori, Riccarda and Calabrese, Pasquale},
  journal = {Phys. Rev. B},
  volume = {103},
  issue = {4},
  pages = {L041104},
  numpages = {7},
  year = {2021},
  publisher = {American Physical Society},
  doi = {10.1103/PhysRevB.103.L041104},
  url = {https://link.aps.org/doi/10.1103/PhysRevB.103.L041104}
}

@article{parez2021exact,
   title={Exact quench dynamics of symmetry resolved entanglement in a free fermion chain},
   ISSN={1742-5468},
   url={http://dx.doi.org/10.1088/1742-5468/ac21d7},
   DOI={10.1088/1742-5468/ac21d7},
   number={9},
   journal={J. Stat. Mech.},
   publisher={IOP Publishing},
   author={Parez, Gilles and Bonsignori, Riccarda and Calabrese, Pasquale},
   year={2021},
 pages={093102} }

@article{rath2023entanglement,
  title = {{Entanglement Barrier and its Symmetry Resolution: Theory and Experimental Observation}},
  author = {Rath, Aniket and Vitale, Vittorio and Murciano, Sara and Votto, Matteo and Dubail, J\'er\^ome and Kueng, Richard and Branciard, Cyril and Calabrese, Pasquale and Vermersch, Beno\^{\i}t},
  journal = {PRX Quantum},
  volume = {4},
  issue = {1},
  pages = {010318},
  numpages = {39},
  year = {2023},
  publisher = {American Physical Society},
  doi = {10.1103/PRXQuantum.4.010318},
  url = {https://link.aps.org/doi/10.1103/PRXQuantum.4.010318}
}

@article{ares2023entanglement,
   title={Entanglement asymmetry as a probe of symmetry breaking},
   volume={14},
   ISSN={2041-1723},
   url={http://dx.doi.org/10.1038/s41467-023-37747-8},
   DOI={10.1038/s41467-023-37747-8},
   number={1},
   journal={Nature Commun.},
   publisher={Springer Science and Business Media LLC},
   author={Ares, Filiberto and Murciano, Sara and Calabrese, Pasquale},
   year={2023},
 pages={2036} }

@article{ESSLER2023127572,
title = {{A short introduction to Generalized Hydrodynamics}},
journal = {Physica A},
volume = {631},
pages = {127572},
year = {2023},
issn = {0378-4371},
doi = {https://doi.org/10.1016/j.physa.2022.127572},
url = {https://www.sciencedirect.com/science/article/pii/S0378437122003971},
author = {Fabian H.L. Essler}
}

@article{Zamolodchikov:1992zr,
    author = "Zamolodchikov, Alexander B. and Zamolodchikov, Alexei B.",
    title = "{Massless factorized scattering and sigma models with topological terms}",
    doi = "10.1016/0550-3213(92)90136-Y",
    journal = "Nucl. Phys. B",
    volume = "379",
    pages = "602",
    year = "1992"
}

@Article{Bazhanov1996,
author={Bazhanov, Vladimir V.
and Lukyanov, Sergei L.
and Zamolodchikov, Alexander B.},
title={Integrable structure of conformal field theory, quantum {KdV} theory and Thermodynamic {B}ethe Ansatz},
journal={Commun.  Math. Phys.},
year={1996},
day={01},
volume={177},
number={2},
pages={381},
issn={1432-0916},
doi={10.1007/BF02101898},
url={https://doi.org/10.1007/BF02101898}
}

@article{CASINI2004142,
title = {A finite entanglement entropy and the c-theorem},
journal = {Phys. Lett. B},
volume = {600},
number = {1},
pages = {142},
year = {2004},
issn = {0370-2693},
doi = {https://doi.org/10.1016/j.physletb.2004.08.072},
url = {https://www.sciencedirect.com/science/article/pii/S037026930401264X},
author = {H. Casini and M. Huerta},
}

@article{Casini_2007,
doi = {10.1088/1751-8113/40/25/S57},
url = {https://doi.org/10.1088/1751-8113/40/25/S57},
year = {2007},
publisher = {},
volume = {40},
number = {25},
pages = {7031},
author = {Casini, H and Huerta, M},
title = {A c-theorem for entanglement entropy},
journal = {J. Phys. A},
}

@article{groha2018full,
   title={{Full counting statistics in the transverse field Ising chain}},
   volume={4},
   ISSN={2542-4653},
   url={http://dx.doi.org/10.21468/SciPostPhys.4.6.043},
   DOI={10.21468/scipostphys.4.6.043},
   number={6},
   journal={SciPost Phys.},
   publisher={Stichting SciPost},
   author={Groha, Stefan and Essler, Fabian and Calabrese, Pasquale},
   year={2018},
    pages=043}

@article{myers2020,
	title={{Transport fluctuations in integrable models out of equilibrium}},
	author={Jason Myers and M. J. Bhaseen and Rosemary J. Harris and Benjamin Doyon},
	journal={SciPost Phys.},
	volume={8},
	pages={007},
	year={2020},
	publisher={SciPost},
	doi={10.21468/SciPostPhys.8.1.007},
	url={https://scipost.org/10.21468/SciPostPhys.8.1.007},
}

@article{murciano2022negativity,
  title = {{Negativity Hamiltonian: An Operator Characterization of Mixed-State Entanglement}},
  author = {Murciano, Sara and Vitale, Vittorio and Dalmonte, Marcello and Calabrese, Pasquale},
  journal = {Phys. Rev. Lett.},
  volume = {128},
  issue = {14},
  pages = {140502},
  numpages = {7},
  year = {2022},
  publisher = {American Physical Society},
  doi = {10.1103/PhysRevLett.128.140502},
  url = {https://link.aps.org/doi/10.1103/PhysRevLett.128.140502}
}

@article{NESS,
	doi = {10.1007/s00023-014-0314-8},
	year = 2014,
	publisher = {Springer Science and Business Media {LLC}},
	volume = {16},
	number = {1},
	pages = {113},
	author = {Denis Bernard and Benjamin Doyon},
	title = "{Non-Equilibrium Steady States in Conformal Field Theory}",
	journal = {Ann. Henri Poincar{\'{e}}}
}

@article{NESS2,
doi = {10.1088/1742-5468/2016/03/033104},
url = {https://dx.doi.org/10.1088/1742-5468/2016/03/033104},
year = {2016},

publisher = {IOP Publishing and SISSA},
number = {3},
pages = {033104},
author = {Denis Bernard and Benjamin Doyon},
title = "{A Hydrodynamic Approach to Non-Equilibrium Conformal Field Theories}",
journal = {J. Stat. Mech}
}

@book{Korepin_1993, place={Cambridge}, series={Cambridge Monographs on Mathematical Physics}, title={Quantum Inverse Scattering Method and Correlation Functions}, publisher={Cambridge University Press}, author={Korepin, V. E. and Bogoliubov, N. M. and Izergin, A. G.}, year={1993}, collection={Cambridge Monographs on Mathematical Physics}}

@article{orbach1958,
  title = {Linear Antiferromagnetic Chain with Anisotropic Coupling},
  author = {Orbach, R.},
  journal = {Phys. Rev.},
  volume = {112},
  issue = {2},
  pages = {309},
  numpages = {0},
  year = {1958},
  publisher = {American Physical Society},
  doi = {10.1103/PhysRev.112.309},
  url = {https://link.aps.org/doi/10.1103/PhysRev.112.309}
}

@article{rottoli2023finite,
   title={{Finite temperature negativity Hamiltonians of the massless Dirac fermion}},
   ISSN={1029-8479},
   volume={2023},
   url={http://dx.doi.org/10.1007/JHEP06(2023)139},
   DOI={10.1007/jhep06(2023)139},
   number={6},
   journal={J. High Energy Phys.},
   publisher={Springer Science and Business Media LLC},
   author={Rottoli, Federico and Murciano, Sara and Calabrese, Pasquale},
   year={2023},
pages =139}

@article{Pollmann2013,
  title = {Linear quantum quench in the {Heisenberg XXZ chain: Time-dependent Luttinger-model description of a lattice system}},
  author = {Pollmann, Frank and Haque, Masudul and D\'ora, Bal\'azs},
  journal = {Phys. Rev. B},
  volume = {87},
  issue = {4},
  pages = {041109(R)},
  numpages = {4},
  year = {2013},
  month = {Jan},
  publisher = {American Physical Society},
  doi = {10.1103/PhysRevB.87.041109},
  url = {https://link.aps.org/doi/10.1103/PhysRevB.87.041109}
}

@article{travaglino2026dynamical,
      title={Dynamical correlation functions of extensive charges after global quantum quenches}, 
      author={Riccardo Travaglino and Katja Klobas and Bruno Bertini and Pasquale Calabrese},
      year={2026},
      eprint={2607.19208},
      archivePrefix={arXiv},
}

@article{Latorre_2009,
doi = {10.1088/1751-8113/42/50/504002},
url = {https://doi.org/10.1088/1751-8113/42/50/504002},
year = {2009},

publisher = {},
volume = {42},
number = {50},
pages = {504002},
author = {Latorre, J I and Riera, A},
title = {A short review on entanglement in quantum spin systems},
journal = {J. Phys. A}
}

@article{Ryu2006,
  title = {Holographic Derivation of Entanglement Entropy from the anti--de {Sitter} Space/Conformal Field Theory Correspondence},
  author = {Ryu, Shinsei and Takayanagi, Tadashi},
  journal = {Phys. Rev. Lett.},
  volume = {96},
  issue = {18},
  pages = {181602},
  numpages = {4},
  year = {2006},
  publisher = {American Physical Society},
  doi = {10.1103/PhysRevLett.96.181602},
  url = {https://link.aps.org/doi/10.1103/PhysRevLett.96.181602}
}

@article{Mossel_2012,
doi = {10.1088/1751-8113/45/25/255001},
url = {https://doi.org/10.1088/1751-8113/45/25/255001},
year = {2012},
publisher = {IOP Publishing},
volume = {45},
number = {25},
pages = {255001},
author = {Mossel, Jorn and Caux, Jean-Sébastien},
title = {Generalized {TBA and generalized Gibbs}},
journal = {J. Phys. A}
}

@article{Arias2023,
author={Arias, Ra{\'u}l
and Di Giulio, Giuseppe
and Keski-Vakkuri, Esko
and Tonni, Erik},
title={Probing {RG} flows, symmetry resolution and quench dynamics through the capacity of entanglement},
journal={J. High Energy Phys.},
year={2023},
volume={03},
pages={175},
issn={1029-8479},
doi={10.1007/JHEP03(2023)175},
url={https://doi.org/10.1007/JHEP03(2023)175}
}

@article{arias2023b,
  title = {Sequences of resource monotones from modular Hamiltonian polynomials},
  author = {Arias, Ra\'ul and de Boer, Jan and Di Giulio, Giuseppe and Keski-Vakkuri, Esko and Tonni, Erik},
  journal = {Phys. Rev. Res.},
  volume = {5},
  issue = {4},
  pages = {043082},
  numpages = {26},
  year = {2023},
  publisher = {American Physical Society},
  doi = {10.1103/PhysRevResearch.5.043082},
  url = {https://link.aps.org/doi/10.1103/PhysRevResearch.5.043082}
}

@article{Giudici_2018,
   title={Entanglement {Hamiltonians} of lattice models via the {Bisognano-Wichmann} theorem},
   volume={98},
   pages={134403},
   ISSN={2469-9969},
   url={http://dx.doi.org/10.1103/PhysRevB.98.134403},
   DOI={10.1103/physrevb.98.134403},
   number={13},
   journal={Phys. Rev. B},
   publisher={American Physical Society (APS)},
   author={Giudici, G. and Mendes-Santos, T. and Calabrese, P. and Dalmonte, M.},
   year={2018} }

@article{Ryu_2006_aspects,
doi = {10.1088/1126-6708/2006/08/045},
url = {https://doi.org/10.1088/1126-6708/2006/08/045},
year = {2006},
publisher = {},
volume = {08},
pages = {045},
author = {Shinsei Ryu and Tadashi Takayanagi},
title = {Aspects of holographic entanglement entropy},
journal = {J. High Energy Phys.},
}

@article{borsi2020currents,
  title = {Current Operators in {Bethe} Ansatz and Generalized Hydrodynamics: An Exact Quantum-Classical Correspondence},
  author = {Borsi, M\'arton and Pozsgay, Bal\'azs and Pristy\'ak, Levente},
  journal = {Phys. Rev. X},
  volume = {10},
  issue = {1},
  pages = {011054},
  numpages = {26},
  year = {2020},
  
  publisher = {American Physical Society},
  doi = {10.1103/PhysRevX.10.011054},
  url = {https://link.aps.org/doi/10.1103/PhysRevX.10.011054}
}

@article{Borsi_2021,
doi = {10.1088/1742-5468/ac0f6b},
url = {https://doi.org/10.1088/1742-5468/ac0f6b},
year = {2021},
publisher = {IOP Publishing and SISSA},
number = {9},
pages = {094001},
author = {Borsi, Márton and Pozsgay, Balázs and Pristyák, Levente},
title = {Current operators in integrable models: a review},
journal = {J. Stat. Mech.}}

@article{deboer2019,
  title = {Aspects of capacity of entanglement},
  author = {de Boer, Jan and J\"arvel\"a, Jarkko and Keski-Vakkuri, Esko},
  journal = {Phys. Rev. D},
  volume = {99},
  issue = {6},
  pages = {066012},
  numpages = {34},
  year = {2019},

  publisher = {American Physical Society},
  doi = {10.1103/PhysRevD.99.066012},
  url = {https://link.aps.org/doi/10.1103/PhysRevD.99.066012}
}

@article{Laflorencie_2016,
   title={Quantum entanglement in condensed matter systems},
   volume={646},
   ISSN={0370-1573},
   url={http://dx.doi.org/10.1016/j.physrep.2016.06.008},
   DOI={10.1016/j.physrep.2016.06.008},
   journal={Phys. Rep.},
   publisher={Elsevier BV},
   author={Laflorencie, Nicolas},
   year={2016},
   pages={1} }

@article{headrick2019lectures,
      title={Lectures on entanglement entropy in field theory and holography}, 
      author={Matthew Headrick},
      year={2019},
      eprint={1907.08126},
      archivePrefix={arXiv},
      primaryClass={hep-th},

}

@Article{calabrese_ln,
	title={{Entanglement spreading in non-equilibrium integrable systems}},
	author={Pasquale Calabrese},
	journal={SciPost Phys. Lect. Notes},
	pages={20},
	year={2020},
	publisher={SciPost},
	doi={10.21468/SciPostPhysLectNotes.20},
	url={https://scipost.org/10.21468/SciPostPhysLectNotes.20},
}

@article{bethezur1931,
    title = {Zur {Theorie} der {Metalle}: {I}. {Eigenwerte} und {Eigenfunktionen} der linearen {Atomkette}},
    volume = {71},
    copyright = {http://www.springer.com/tdm},
    issn = {1434-6001, 1434-601X},
    shorttitle = {Zur {Theorie} der {Metalle}},
    url = {http://link.springer.com/10.1007/BF01341708},
    doi = {10.1007/BF01341708},
    number = {3-4},
    urldate = {2026-05-25},
    journal = {Zeits. Phys.},
    author = {Bethe, H.},
    year = {1931},
    pages = {205},
}

@article{Alba_Renyi_2017,
   title={Rényi entropies after releasing the {N}éel state in the {XXZ} spin-chain},
   ISSN={1742-5468},
   url={http://dx.doi.org/10.1088/1742-5468/aa934c},
   DOI={10.1088/1742-5468/aa934c},
   number={11},
   journal={J. Stat. Mech.},
   publisher={IOP Publishing},
   author={Alba, Vincenzo and Calabrese, Pasquale},
   year={2017},
    pages={113105} }
\end{document}